\documentstyle[psfig]{aa}

\def\magcir{\raise -2.truept\hbox{\rlap{\hbox{$\sim$}}\raise5.truept
\hbox{$>$}\ }}

   \title{A BeppoSAX observation of MKN6}

   \author{A.~Malizia\inst{1}, L.~Bassani\inst{1}, M.~Capalbi\inst{2}, A.~C.~Fabian\inst{3}, 
F.~Fiore\inst{4}, F.~Nicastro\inst{5}}

   \offprints{A. Malizia (malizia@bo.iasf.cnr.it)}

   \institute{IASF/CNR, via Piero Gobetti 101, I-40129 Bologna, Italy \and ASI Science Data Center, c/o 
  ESA-ESRIN, Via Galileo Galilei, I-00044 Frascati, Italy \and Institute of Astronomy, University 
  of Cambridge, Madingley Road, Cambridge, CB3 0HA \and INAF, Osservatorio Astronomico di Roma, 
  Via dell'Osservatorio, I-00044 Monteporzio Catone, Italy}
    \date{Received / accepted}

   \titlerunning{A BeppoSAX observation of MKN6}

   \authorrunning{}

\begin{document}

\abstract{We have used the BeppoSAX satellite to study the broad band (0.5-100 keV)
X-ray spectrum of the Seyfert 1.5 galaxy MKN6. The source is
characterized by a power law of $\Gamma$=1.7$^{+0.08}_{-0.07}$ and
there is no strong evidence for either a reflection bump or a high
energy cut-off.  We have detected a narrow iron line at 6.4 keV (rest
frame) with an equivalent width of 98$^{+33}_{-35}$ eV. MKN6 also
exhibits strong and complex absorption.  At least two components
(N$_{H_{1}}$=1.34$^{+0.4}_{-0.4}$ $\times$ 10$^{22}$ cm$^{-2}$ and
N$_{H_{2}}$=4.18$^{+2.2}_{-1.3}$ $\times$ 10$^{22}$ cm$^{-2}$) are
present and they both partially cover the source with covering fractions
of $\sim$90\% and $\sim$50\% respectively.  Comparison with a previous
ASCA observation indicates that in both absorbing columns the N$_{H}$
is variable over a 2 year timescale, while the covering fractions are
constant over the same amount of time. The state of each absorber is
cold or mildly photoionized.  The Broad Line Region (BLR) is suggested
as the possible location for this complex absorption.

}
 \maketitle


\section{Introduction}
MKN6 (IC 450) is a Seyfert 1.5 galaxy which has not been observed
often in the X-ray regime despite its brightness.  The only X-ray
study of MKN6 is that of ASCA (Feldmeier et al. 1999) which revealed a
very interesting spectrum in the 0.6-9.5 keV band, showing heavy and
complex intrinsic absorption in the AGN nucleus, a power law continuum
of $\Gamma \sim$1.6 and an apparently broad 6.4 keV iron K$\alpha$
line.  On the other hand MKN6 is the best known example of an NGC4151
analogue: both are Seyfert 1.5 galaxies characterized by ionization
cones, strong and complex X-ray absorption and a flat high energy
spectrum.  All these features are quite atypical of type 1 objects and
more commonly found in Seyfert 2s.  Observationally Seyfert 1.5s are
widely recognized as a distinct class of Seyfert galaxies
(Osterbrock \& Koski 1976, Cohen 1983) although their nature has not
yet been fully understood.  According to Nagao, Taniguchi \& Murayama
(2000), it could be that they are seen from an intermediate viewing angle between
Seyfert 1s and Seyfert 2s so that a significant part of the broad line
region (BLR) is obscured by a dusty torus.  Alternatively they may be
Seyfert 1s where a large part of the BLR emission is obscured by
dense, clumpy gas clouds.  Finally, they could be Seyfert 2 galaxies
where part of the BLR emission can be seen through some optically thin
regions of the dusty torus.  The latter idea is consistent with the
model suggested by Feldmeier et al.  (1999) to explain the heavy
absorption measured in MKN6 by ASCA i.e. a dusty torus with a low
density atmosphere.  In this scenario our line of sight skims the
surface of the torus and thus passes through the atmosphere which is
assumed to have a low column density of the order of
10$^{20}$-10$^{21}$ cm$^{-2}$ of neutral or low ionization material.
However the absorption measured by ASCA in MKN6 (Feldmeier et
al. 1999) is much stronger than that expected in a torus atmosphere
(i.e. $\sim$ 10$^{23}$ cm$^{-2}$) and over an order of magnitude
higher than that expected based on observations at longer wavelengths.
This inconsistency was overcome by Feldmeier et al. (1999) 
by assuming either an additional absorber located inside the torus atmosphere or
that the atmosphere is composed of relatively dust-free gas.\\ In
order to assess the true slope and the broad band X-ray spectrum of
MKN6, to compare its characteristics with those of NGC4151 and to understand
the nature of these two peculiar Seyferts, we performed an observation
over the 0.5-100 keV band with the BeppoSAX satellite.

\section{Spectral analysis}
\subsection{The BeppoSAX Observation and Data Reduction}
MKN6 was targeted by the BeppoSAX NFI in 1999 from September 14th to September
17th.  The effective exposure times were 48 ks for the LECS, 109
ks for the MECS23 and 52 ks for the PDS.  Spectra  were extracted
from a region centered on MKN6 with a radius of 4 arcmin.  LECS and
MECS background spectra were extracted from blank sky fields using
regions of the same size in detector coordinates.  The net count rate
was $5.02\pm0.1\times10^{-2}$ cts/s in the (0.5-4 keV) LECS energy
band, $0.25\pm0.002$ cts/s in the (2-10 keV) MECS energy range, and
$0.5\pm0.02$ cts/s in the (20-100 keV) PDS band.  Source plus
background light curves did not show any significant variation,
therefore a timing analysis has not been performed and all data have
been grouped for spectral analysis. \\ Standard data reduction was
performed using the software package "SAXDAS" (see
http://www.sdc.asi.it/software and the Cookbook for BeppoSAX NFI
spectral analysis, Fiore, Guainazzi \& Grandi 1998).  Spectral fits
were performed using the XSPEC 11.0.1 software package and public
response matrices as from the 1998 November issue.  
The spectra were rebinned in order to have at least 20 counts per channel,
this allows the use of the $\chi^2$
method in determining the best fit parameters, since the distribution
in each channel can be considered Gaussian.  Constant factors have
been introduced in the fitting models in order to take into account
the inter-calibration systematic uncertainties between instruments
(Fiore, Guainazzi \& Grandi 1998).  All the quoted errors correspond
to 90\% confidence level for one interesting parameter ($\Delta\chi^2$
= 2.71).  All the models used in what follows contain an additional
term to allow for the absorption of X-rays due to our galaxy, which in
the direction of MKN6 is N$_{H_{Gal}}$ = 6.4 $\times$ 10$^{20}$ atoms
cm$^{-2}$ (based on 21-cm radio observations, provided by
XSPEC).

\begin{figure}
\psfig{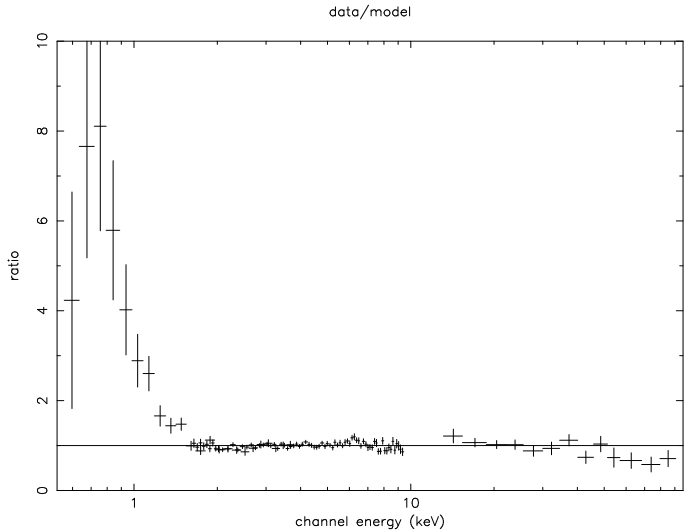}
\caption{Data to power law plus Galactic absorption model ratio.}
\end{figure}

\subsection{Data Analysis}
The broad band (0.5-100 keV) spectrum of MKN6 was first modeled with
an absorbed power law.  This model provides a flat photon index
($\Gamma$ $\approx$ 1.5) and an N$_{H}$ $\approx$ 2 $\times$ 10$^{22}$
cm$^{-2}$, but gives a reduced $\chi^{2}$ of 228 for 105 degrees of
freedom and hence is unacceptable. In fact as shown in figure 1, it leaves systematic
residuals throughout the whole spectrum consisting of:
(a) an evident excess emission at low energies; (b) a
marginal line-like feature at $\sim$ 6.4 keV; and (c) possibly a high
energy cut-off.  Adding a gaussian line to this model (model 1 in
table 1), gives a $\Delta\chi^{2}$ of 33 for three more parameters, so
that the line is statistically significant at $>$99\% confidence 
using the F-test.  The line is centered at 6.39$^{+0.13}_{-0.11}$ keV
(rest frame), has an equivalent width EW=114$^{+14}_{-70}$ eV and is
consistent with being narrow at the 99\% confidence level as
demonstrated in figure 2; therefore in the following we have fixed the
value of sigma to zero.  The resulting flat photon index
($\Gamma$=1.52$\pm0.03$) and the count excess at soft energies could
indicate the presence of more intrinsic absorption than assumed in the
first instance.  We have therefore added to the neutral absorber a
partial covering absorption model ({\sc PCFABS}, model 2) which has
been successfully applied to this object in the ASCA observation
(Feldmeier et al. 1999).  The addition of this partial covering
absorber improves the quality of the fit at greater than $>$99\%
significance level ($\Delta\chi^{2}$=103 for two more parameters).
Using this model, the source is completely covered by cold uniform
material with N$_{H}\sim$4-9 $\times$ 10$^{21}$ atoms cm$^{-2}$, and
partially covered ($\sim$ 70\%) by another with N$_{H}\sim$3 $\times$
10$^{22}$ atoms cm$^{-2}$. Despite the goodness of this fit some
residuals below 1 keV are still present. \\
We have investigated if the presence of a diffuse hot gas located outside 
the intrinsic absorption and  heated by starburst activity in the 
host galaxy, could be responsible for the soft excess component.
The addition of a thermal component (Raymond-Smith model in XSPEC)
to model 2 does not improve the quality of the fit ($\chi^{2}$=203 with 99 d.o.f).
Further, this model was also rejected by Feldmeier et al. (1999) when applied to the ASCA data.\\
If we substitute in model 2 the neutral absorber with another partial covering one (model 3), the fit improves
further ($>$99\%, $\Delta\chi^{2}$=8 with one more parameter) and
provides our best fit model to the data as also found with the ASCA
data.  Since the width of the line is sensitive to the underlying
continuum, we have left the width free to vary in model 3 and also in
this case the line is consistent with being narrow. This is also true
for any of the models listed in table 1.  In figure 3 the broad band
(0.5-100 keV) spectrum of MKN6 fitted with model 3 is shown. \\
An alternative model used to fit a multicolumn absorber is
the {\it dual-absorber}.
This model requires different assumptions on the source + absorber
geometry and thus a different interpretation of the data.  In the
dual-absorber model, two different columns cover the source and the
relative normalization of the power law absorbed by these two columns
gives the percentage of the source covered by each N$_{H}$ in such a
way that the sum of both percentages is 100.  We have applied this model
to our data (model 4) but no improvement in the quality of the fit has
been found.\\ 
To check the robustness of our findings, we have also
investigated the presence of a warm medium instead of a cold one
via the ABSORI model in XSPEC.  In the case when the cold medium is
represented by a uniform absorber the $\chi^{2}$ worsens significantly
(195/101) thus this model must be rejected.  If instead the cold
medium is assumed to be an absorber partially covering the source, a
combination which was successfully used to model the complex
absorption of NGC4151 (Schurch and Warwick 2002, Piro et al. 2002),
then we obtain a statistically acceptable fit to our data.  However we
are not able to constrain N$_{H_{warm}}$ and $\xi$ with this
model. Reasonable values of the ionization parameter, deduced from
previous studies of Seyfert galaxies (Perola et al. 2002) and
compatible with the lack of strong warm absorption features in the
spectrum, all provide acceptable fits: for example assuming a column
density similar to that obtained in the case of uniform covering,
N$_{H_{warm}}$=3 $\times$ 10$^{22}$ cm$^{-2}$ and $\xi$ $\approx$ 10
erg cm s$^{-1}$ gives $\chi^{2}$=109/100.  Therefore we cannot exclude
that the absorbing gas is just mildly ionized on the basis of our data,
although we are not able to constrain its properties significantly.\\
We have also searched for the presence of a high energy cut-off and a
reflection component which are typically observed in Seyfert 1
galaxies (e.g. Perola et al. 2002) by replacing the power law with the
PEXRAV model (model 5).  However, we do not find strong evidence for
either of these two components and we are only able to put upper
limits on the parameters values: R$<$1.2 and E$_{c}>$70 keV.

\begin{figure}
\psfig{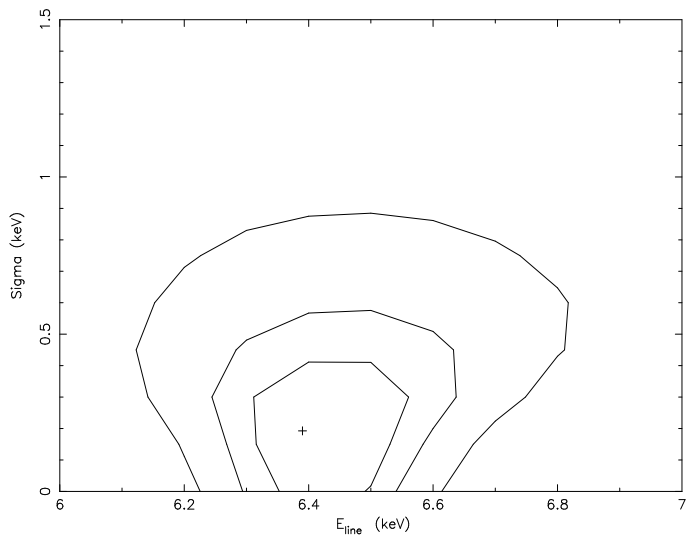}
\caption{Confidence contours of the Fe K$\alpha$ line versus its width (sigma). 
The line is consistent with being narrow at 99\% confidence level.}
\end{figure}

\begin{figure}
\psfig{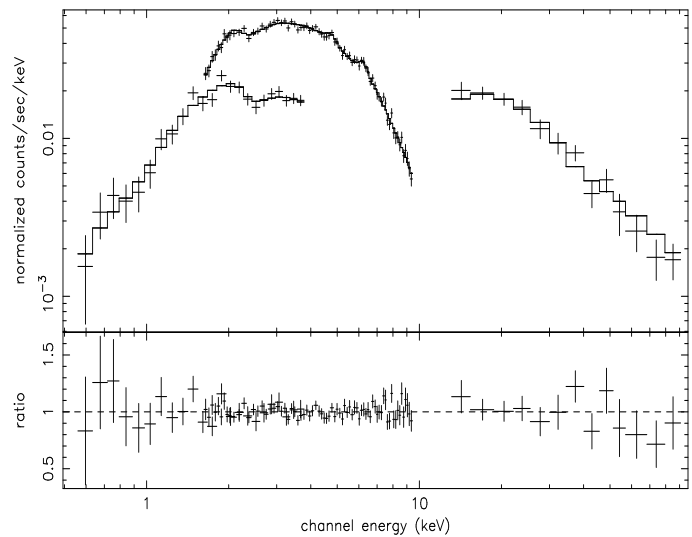}
\caption{Broad band (0.5-100 keV) spectrum of MKN6 fitted with model 3 of table 1.}
\end{figure}

\subsection{Comparison with the ASCA results}
In this subsection we briefly compare two X-ray studies of MKN6: the
ASCA observation performed on April 1997 and our BeppoSAX measurement
from September 1999.  The source shows flux variability in the 2-10
keV energy range; in fact the ASCA absorbed (unabsorbed) flux is
$\sim$1 (1.4) $\times$ 10$^{-11}$ erg cm$^{-2}$ s$^{-1}$ while for
BeppoSAX it is $\sim$ 2.5 (3.1) $\times$ 10$^{-11}$ erg cm$^{-2}$
s$^{-1}$.  These flux values are obtained using the same model (model
3 in table 1) while the absorption correction is performed by forcing
both column densities to zero.  \\
The BeppoSAX observation of MKN6
basically confirms the ASCA results i.e. the detection of heavy and
complex X-ray absorption towards the source nucleus.  In both
observations a double partial covering model provides the best
statistical fit to the data but from the comparison of our results
with those of ASCA a variation in the column densities has been
detected.  Our analysis of ASCA data provides
N$_{H_{1}}$ = 2.99$^{+0.13}_{-0.21}$ $\times$ 10$^{22}$ cm$^{-2}$ with
a covering fraction of $\sim$90\% and N$_{H_{2}}$ = 11$^{+3.1}_{-2.2}$
$\times$ 10$^{22}$ cm$^{-2}$ with a covering fraction of $\sim$50\%.
These values are consistent within the errors with those reported by
Feldmeier et al. (1999).  BeppoSAX detected smaller column densities
of 1.2 and 3.6 $\times$ 10$^{22}$ cm$^{-2}$ with covering fractions
C$_{f_{1}}$=93$^{+0.3}_{-0.7}$\% and C$_{f_{2}}$=52$^{+1.7}_{-1.6}$\%
respectively.  Therefore, while the covering fractions C$_{f_{1}}$ and
C$_{f_{2}}$ are not changed with respect to the ASCA observation,
within their individual statistical uncertainties, a significant
variation in the column densities of the absorbing materials is found
over a two year timescale.  
This is evident in figure 4 where
the LECS+MECS BeppoSAX data are fitted with model 3 but imposing the best
fit parameters obtained from ASCA: clearly the two data sets cannot be
reconciled without changes in the spectral shape parameters.  To
further prove absorption variations, we show in figure 5 the
confidence contours of both column densities in the BeppoSAX (solid
lines) and ASCA (dashed line) observations.  Since both covering
fractions are very similar and relatively well constrained, we have
fixed them to their best fit values.
A check for variations in the absorption over a shorter time-scale has also been
performed. Since the BeppoSAX observation does not show any significant variation
in the continuum light curve, the whole observation (about 100 ks MECS exposure) has been divided 
in two roughly equal parts and the same spectral analysis of the entire observation has been 
repeated on both segments of the observation.
We find no significant variation in any of the spectral parameters, 
and in particular in the amount of absorption. We can therefore 
conclude that no short timescale (one day) variations are present.\\
Regarding the iron line, we note that the broadening seen by ASCA
($\sigma \sim$230 eV) is still consistent with our measurement of a narrow feature
(see figure 2).\\ 
Finally, we measure a slightly harder spectrum ($\Gamma$=1.7$^{+0.08}_{-0.07}$)
than 1.6 as fixed in the ASCA data.

\begin{table*}
\begin{center}
\caption{MKN6: Spectral Analysis}
\begin{tabular}{clclclllccl}
\hline
Model & N$_{H_{1}} \times$ 10$^{22}$ & C$_{f_{1}}$ & N$_{H_{2}}
\times$ 10$^{22}$ & C$_{f_{2}}$ & $\Gamma$ & E$_{Line}$ (keV) & EW
(eV) & R & E$_c$ (keV) & $\chi^{2}$/$\nu$ \\
\hline
 (1) & 1.94$^{+0.09}_{-0.14}$ & - & - & - & 1.52$^{+0.03}_{-0.03}$ &
 6.39$^{+0.13}_{-0.11}$ & 114$^{+14}_{-70}$ & - & - & 195/103 \\ 
(2) & 0.61$^{+0.26}_{-0.25}$ & - & 3.02$^{+0.86}_{-0.51}$ &
 0.75$^{+0.09}_{-0.11}$ & 1.66$^{+0.07}_{-0.06}$ &
 6.40$^{+0.06}_{-0.12}$ & 108$^{+36}_{-48}$ & - & - & 92/101 \\ 
(3) & 1.34$^{+0.44}_{-0.42}$ & 0.93$^{+0.03}_{-0.07}$ &
 4.18$^{+2.23}_{-1.29}$ & 0.52$^{+0.17}_{-0.16}$ &
 1.70$^{+0.08}_{-0.07}$ & 6.42$^{+0.15}_{-0.14}$ & 97$^{+43}_{-44}$ &
 - & - & 84/100 \\ 
(4) & 0.54$^{+0.34}_{-0.21}$ & 0.30$^{+0.22}_{-0.13}$ & 3.05$^{+1.06}_{-0.40}$ &
 0.70$^{+0.13}_{-0.22}$ & 1.66$^{+0.09}_{-0.05}$ &
 6.42$^{+0.14}_{-0.14}$ & 137$^{+52}_{-54}$ & - & - & 92/101 \\ 
(5) & 1.25$^{+0.47}_{-0.51}$ & 0.93$^{+0.02}_{-0.06}$ &
 3.57$^{+1.84}_{-1.07}$ & 0.53$^{+0.20}_{-0.19}$ &
 1.63$^{+0.08}_{-0.06}$ & 6.42$^{+0.04}_{-0.06}$ & 93$^{+34}_{-34}$ &
 $<$1.2 & $>$ 70 & 83/98 \\
\hline
\end{tabular}
\end{center}
\end{table*}

\begin{figure}
\psfig{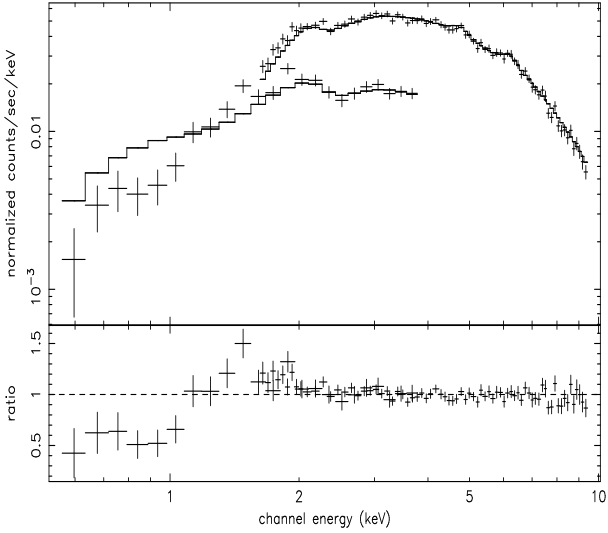}
\caption{LECS+MECS BeppoSAX data fitted with model 3 but imposing the best fit parameters obtained by ASCA
for both partial covering absorbers.}
\end{figure}

\begin{figure}
\psfig{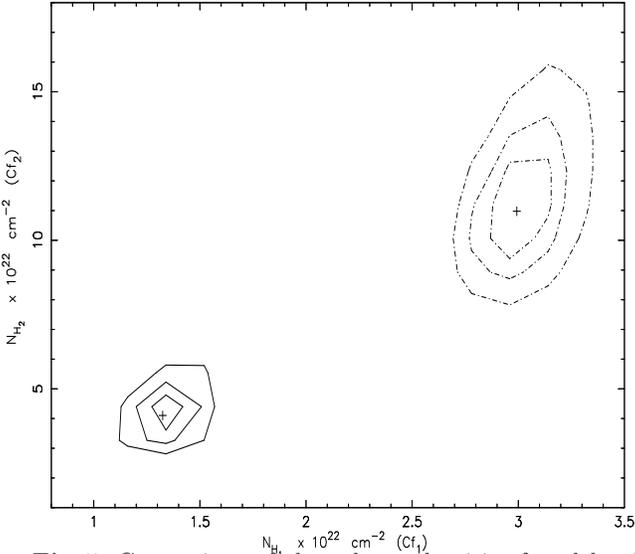}
\caption{Comparison of the column densities found by ASCA (dashed lines) and BeppoSAX (solid lines).}
\end{figure}

\section{Discussion and Conclusions}
The MKN6 spectrum observed by BeppoSAX resembles that of NGC4151 in
the absorption properties although the broad band spectral
characteristics (the power law, reflection component and high energy
cut-off) are similar to more classical Seyfert 1 objects (Perola et
al. 2002).  The absorption is complex and variable.  At least two
components, each with its own column density and covering fraction, are
present. In both columns the N$_{H}$ is variable over a two year
timescales while the covering is constant over the same time. The
state of each absorber is cold or mildly photoionized, in fact a
combination of cold and warm gas as seen in NGC4151 is not ruled out
by the data.  Recently Maiolino et al. (2001) found that the
E(B-V)/N$_{H_{X}}$ ratio estimated from optical/infrared broad band
lines and X-ray data is lower than Galactic in a sample of AGNs,
including MKN6; this issue was discussed in detail also by Feldmeier
at al (1999).  Our results of a variable N$_{H_{X}}$, clearly suggest
caution in adopting such a ratio as an indicator of anomalous
absorption properties. Furthermore, the optical band lines in MKN6 are
known to be variable in time (Sergeev et al. 1999); therefore an
estimate of the E(B-V)/N$_{H}$ ratio strongly requires simultaneous
optical and X-ray measurements.\\ The variation observed in N$_{H}$
can be explained either by a change in the ionization state of each
absorber due to a change in the incoming radiation and/or to variation
in the amount of the absorbing gas along the line of sight.\\ 
A joint fit of ASCA/GIS and BeppoSAX/MECS data are compatible with a
change in the $\xi$ parameter although the limitation of the ABSORI
model in XSPEC does not allow us to firmly prove a variation in the
ionization state of the absorption.
The other possibility is that the amount of gas along the line of
sight varied between the ASCA and BeppoSAX observation.
In this case the absorber must be clumpy and the variation timescales are
related to the typical crossing time of an absorbing cloud along the line of sight. 
Following the reasoning of Risaliti et al. (2002)  
the location of the absorber is related to the black hole mass, the column density
of the absorber, the cloud density and the absorption variation:
the BeppoSAX data indicate that the  variability timescales are greater  
than 1 day and smaller than 2.4 years. For reasonable values of the other relevant
parameters we find that  the molecular torus 
and the BLR are both viable locations for the
absorption (Risaliti et al. 2002).
In the first case we go back to the original suggestion
of Feldmeier et al. (1999) who propose the atmosphere above the torus
as the X-ray absorber.  However the column densities measured both by
ASCA and BeppoSAX are substantially larger than expected in a torus
atmosphere ($\sim$ 10$^{20}$ - 10$^{21}$ cm$^{-2}$, Wilson 1996).
Considering these difficulties, it is reasonable to consider
the BLR as an alternative scenario.  The BLR location would be
consistent with our variability constraints as well as with the value
of column densities and covering fractions observed. Also the
requirement for a dust free gas, indicated by the greater absorptions
seen in X-rays than at longer wavelengths, is compatible with
the BLR.  Gas lying here is quite likely to be dust free since this
region is within the sublimation radius of the central engine.\\
Furthermore, the partial covering model itself requires that the
location of the absorber is close to, and compatible in size, with the
emitting region thus making the BLR a likely site (Reichert et
al. 1986).  However in this scenario, given the complexity of the
absorber, one must assume that the BLR itself is either obscured by dense
clumpy gas clouds or is itself complex.  This last suggestion is in
line with reverberation studies which show that the BLR is slightly
stratified into multiple emitting zones with strong gradients both in
the ionization parameter and/or density (Baldwin et al. 2003).  It is
therefore likely that changes in the central engine produce variations
in one or more of these stratifications.  It remains to be understood
why such heavy and complex X-ray absorption is only present in a few
objects and not in all Seyfert 1s. The fact that both NGC4151 and MKN6
are Seyfert 1.5 may bear some relation to the peculiarity in their
absorption properties.  Based on the results of our analysis it is
likely that this enigmatic class is made of Seyfert 1's with a BLR
either obscured by extra clumpy gas clouds or peculiar in its
geometry.

\begin{acknowledgements}
This research has made use of SAXDAS linearized and cleaned event
files produced at the BeppoSAX Science Data Center.  This research has
been partially supported by ASI contract I/R/041/02.
\end{acknowledgements}

\end{document}